\documentclass[showpacs,aps,graphicx,twocolumn]{revtex4}%
\usepackage{graphicx}

\begin{document}
\title{Addendum to "Quantum secret sharing between multiparty and multiparty without entanglement" }
\author{Fu-Guo Deng,$^{1,2,}$ Feng-Li Yan,$^{3,4}$  Xi-Han Li,$^{1}$ Chun-Yan Li,$^{1}$ Hong-Yu
Zhou$^{1,2}$ and Ting Gao $^{4,5}$  }
\address{$^1$ The Key Laboratory of Beam Technology and Material Modification of Ministry of
 Education, and Institute of Low Energy
Nuclear Physics, Beijing Normal University, Beijing 100875, China\\
$^2$ Beijing Radiation Center, Beijing 100875,  China\\
$^3$ College of Physics Science and Information Engineering, Hebei Normal University, Shijiazhuang
050016, China\\
$^4$ CCAST (World Laboratory), P.O. Box 8730, Beijing 100080, China\\
$^5$ College of Mathematics and Information Science, Hebei Normal University,
 Shijiazhuang 050016, China\\
}
\date{\today }

\begin{abstract}
Recently, Yan and Gao proposed a quantum secret sharing protocol
between multiparty ($m$ members in group 1) and multiparty ($n$
members in group 2) using a sequence of single photons (Phys. Rev.
A \textbf{72}, 012304 (2005)). We find that it is secure if the
quantum signal transmitted is only a single photon but insecure
with a multi-photon signal as some agents can get the information
about the others' message if they attack the communication with a
Trojan horse. However, security against this attack can be
attained with a simple modification.

\end{abstract}
 \maketitle

In general, secret sharing \cite{Blakley} is used to split a
message ($M_A$) into several pieces which are distributed to
several agents. When the agents collaborate, they can obtain the
message, otherwise they can get nothing.  Quantum secret sharing
(QSS) is the generalization of classical secret sharing into
quantum scenario and becomes an important branch of quantum
communication. It provides a secure way  for not only creating a
private key among several users
\cite{HBB99,KKI,guoqss,longqss,deng2005,deng20052} but also
splitting a piece of classical secret message or quantum
information (an unknown quantum state)
\cite{cleve,Bandyopadhyay,Peng,Karimipour,zhanglm,improving,dengmQSTS,dengpra,Gottesman}.
It has  progressed quickly since Hillery, Bu\v{z}ek and Berthiaume
\cite{HBB99} proposed the original QSS protocol using a
three-photon entangled Greenberger-Horne-Zeilinger  states, and
attracts a lot of attentions in recent years
\cite{HBB99,KKI,deng2005,deng20052,longqss,guoqss,zhangPLA,MZ,Bandyopadhyay,Karimipour,
zhanglm,improving,cleve,Peng,dengpra,dengmQSTS,Gottesman,YanGao,TZG,AMLance}.

Recently, Yan and Gao \cite{YanGao} presented a novel concept for
quantum secret sharing of classical information (a private key),
i.e., QSS between group 1 and group 2 with polarized single
photons. It is secure if the quantum signal transmitted is only a
single photon but insecure with a multi-photon signal as some
agents can get the information about the others' message if they
attack the communication with a Trojan horse.  In this paper, we
will discuss this attack and improve the security of  Yan-Gao QSS
protocol against this attack with a little of modification.

The basic idea of  Yan-Gao QSS protocol \cite{YanGao} can be
described as follows. The members  of group 1 are Alice 1, Alice
2,$...$, and Alice $m$, and those for group 2 are Bob 1, Bob 2,
$...$, and Bob $n$. Group 1 wants to share a private key with
group 2 in such a way that the key can be read out only when all
members in each group cooperate. This task can be completed with
six steps as follows.

(1) Alice 1 generates two random binary strings $A_1$ and $B_1$
whose length is $nN$ bits. She also prepares $nN$ qubits and
divides it into to $n$ pieces. Each piece contains $N$ qubits.
Each qubit is in one of the four states
\begin{eqnarray}
&&\vert \psi_{00}\rangle = \vert 0\rangle,\\
&&\vert \psi_{10}\rangle = \vert 1\rangle,\\
\vert \psi_{01}\rangle && = \vert
+\rangle=\frac{1}{\sqrt{2}}(\vert
0\rangle + \vert 1\rangle),\\
\vert \psi_{11}\rangle && = \vert
-\rangle=\frac{1}{\sqrt{2}}(\vert 0\rangle - \vert 1\rangle),
\end{eqnarray}
where $\vert 0\rangle$ and $\vert 1\rangle$ are the two
eigenvectors of the operator $\sigma_z$ (called it the measuring
basis (MB) $Z$), and $\vert + \rangle$ and $\vert -\rangle$ are
those of the operator $\sigma_x$. That is, Alice 1 prepares the
qubits according to the bit values in $A_1$ and $B_1$. The four
states $\vert 0\rangle$, $\vert 1\rangle$, $\vert +\rangle$, and
$\vert -\rangle$ correspond to the codes 00, 10, 01, and 11,
respectively. The first bit in each code comes from the $A_1$ and
represents the information of the eigenvalue correlated to the
state, and the second bit comes from the $B_1$ and represents the
MB of the state, i.e., the state is the eigenvector of $Z$ if the
second bit is 0, otherwise the state is the eigenvector of $X$.
Alice 1 sends the $nN$ qubits to Alice 2.

(2) Similar to Alice 1, Alice 2 creates two random binary strings
$A_2$ and $B_2$. She operates each of the $nN$ qubits according to
the bits in the two strings in turn, i.e., if the bit in $A_2$ is
0, she performs the unitary operation $I=\vert 0\rangle\langle
0\vert + \vert 1\rangle\langle 1\vert$ on the qubit, otherwise she
performs the unitary operation $U=i\sigma_y=\vert 0\rangle\langle
1\vert - \vert 1\rangle\langle 0\vert$. Simultaneously she has to
operate the qubit according to the bit in $B_2$. That is, if the
bit in $B_2$ is 0, she supplies the unitary operation $I$ on the
qubit, otherwise she performs a Hadamard (\emph{H}) operator on
it. After the two operations on each qubit, Alice 2 sends the $nN$
qubits to Alice 3.

The nice feature of the operation $U$ is that it flips the state
in both measuring bases \cite{deng2005,QOTP,BidQKD}, i.e.,
\begin{eqnarray}
&&U\vert 0\rangle=-\vert 1\rangle, \,\,\,\,\,\,\,\, U\vert 1\rangle=\vert 0\rangle, \\
&&U\vert +\rangle=\vert -\rangle,  \,\,\,\,\,\,\,\, U\vert
-\rangle=-\vert +\rangle.
\end{eqnarray}
The $H$ operation can transfer the states between the two MBs, $Z$
and $X$ \cite{zhanglm,improving}, i.e.,
\begin{eqnarray}
&&H\vert 0\rangle=\vert +\rangle, \,\,\,\,\,\,\,\, H\vert 1\rangle=\vert -\rangle, \\
&&H\vert +\rangle=\vert 0\rangle,  \,\,\,\,\,\,\,\, H\vert
-\rangle=\vert 1\rangle.
\end{eqnarray}
After the two operations done by Alice 2, the photon is randomly
in one of the four states $\{\vert 0\rangle, \vert 1\rangle, \vert
+\rangle, \vert -\rangle\}$, which will prevent Alice 1 from
eavesdropping freely if there is one photon in each signal.

(3) Alice $i$ operates the qubits like Alice 2, $i=3,4,\ldots, m$.

(4) Alice $m$ sends $N$ qubits to each member of group 2, Bob $j$
($j=1, 2, \ldots, n$) in turn. After they receive the qubits, the
members of group 1 publicly announce the strings $B_1, B_2,
\ldots$, and $B_m$, which will reveal the information about the MB
of each qubit operated by Alice $i$ ($i=1, 2, \ldots, m$).

(5) Bob $j$ measures each of his qubits with MB $Z$ or $X$
according to the XOR (i.e., $\oplus$) results of corresponding
bits in the strings $B_1, B_2, \ldots$, and  $B_m$.

(6) All members in group 1 check eavesdropping of this quantum
communication. If the channel is secure, the XOR results of Bob
$j$'s corresponding bits can be used as key bits for secret
sharing.

Yan-Gao protocol may be secure for each member in group 2, say Bob
$j$, as group 1 can detect the eavesdropping done by them, as the
same as BB84 quantum key distribution  protocol \cite{BB84}. But
the eavesdropping done by some members in group 1 is difficult to
be detected, in particular some  agents in group 1 eavesdrop the
communication with a multi-photon signal. We will discuss the
security of Yan-Gao protocol and present a possible improvement of
Yan-Gao protocol security.

The attack done by Alice 1 with a Trojan horse works as follows.
She prepares two photons in the same state and sends them to Alice
2. After operations have been done by the other members in group
1, Alice 1 intercepts the signal and separates a photon from the
signal with a photon number splitter (PNS: 50/50). She sends the
other one in each signal to the members in group 2. After Alice
$i$ ($i=1, 2, \ldots, m$) announces the strings $B_1, B_2,
\ldots$, and $B_m$, Alice 1 can obtain the key freely without the
help of the others in group 1. It means that Yan-Gao protocol is
not secure for Alice 1. Of course, an eavesdropper, Eve (who can
be any one of Alice 2, Alice 3, ..., Alice $m$)  can steal almost
all the information about the unitary operation done by any member
in group 1 with a multi-photon Trojan horse attack except for
Alice 1, as the same as that in Ref. \cite{improving}. That is,
Eve intercepts the original signal and sends Alice $i$ $2^{K+1}$
photons in the same state, say $\phi_0=\vert 0\rangle$. After the
operation is done by Alice $i$, Eve intercepts the photons and
measures with two MBs after splitting it using some PNSs. In
detail, Eve measures half of the photons with MB $Z$, and the
others with MB $X$. The effect of the two operations $\{I, U\}$
according to $A_i$ and  $\{I, H\}$ corresponding to $B_i$ done by
Alice $i$ is equivalent to one of the four operations $\{I, U, H,
\bar{H}\}$ chosen according to the two bits from $A_i$ and $B_i$.
Here $\bar{H}=\frac{1}{\sqrt{2}}(-\vert 0\rangle\langle 0\vert +
\vert 0\rangle\langle 1\vert + \vert 1\rangle\langle 0\vert +
\vert 1\rangle\langle 1\vert)$. The task of the eavesdropping done
by Eve is simplified to distinguish the four unitary operations.
However, they can be distinguished with sufficiently enough
photons. If Alice $i$ performs one of the two operations $\{I,
U\}$ on the fake signal, the outcomes of the measurements on the
$2^K$ photons with MB $Z$ are the same one, otherwise the outcomes
are different. The same result  can be obtained with the
measurements on the other $2^K$ photons with MB $X$ if Alice $i$
performs one of the two operations $\{H, \bar{H}\}$ on the fake
signal. Thus, quantum secret sharing between $m$ members and $n$
members turns into that between $t$ members ($t<m$) and $n$
members. However, since Alice 1 keep $A_1$ in secret and Eve can
not get the information $A_1$ of Alice 1, so two groups still
share a secure key. In a word, Yan-Gao protocol is not secure only
if Alice 1 replaces  the single-photon signal with a multi-photon
one.

In essence, the attack comes from the two facts: one is that the
member in group 1 does not determine whether there is one photon
in each quantum signal received or more; the other is that the
member does not know whether an eavesdropper intercepts the
original signal.

For improving the security of Yan-Gao protocol \cite{YanGao}, a
photon number splitter (PNS: 50/50) and single-photon measurements
are necessary for each of the members in group 1, as the same as
that in Ref. \cite{improving}, except for Alice 1, the one who
prepares the original signal. The measurements with a PNS is shown
in Fig.1. That is, Alice $l$ chooses randomly a sufficiently
enough subset of the $nN$ photons as the samples for eavesdropping
check, and measures each sample with a PNS and two single-photon
detectors after the original quantum signals are transmitted from
Alice $l-1$ to her, see Fig.1.

\begin{figure}[!h]
\begin{center}
\includegraphics[width=6cm,angle=0]{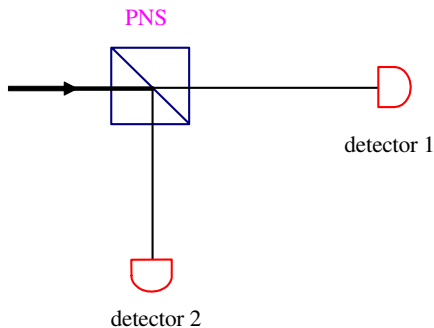} \label{f1}
\caption{ (Color online) The measurements with a photon number
splitter (PNS: 50/50), similar to that in Ref. \cite{improving}.
The members in group 1 choose one of the two MBs, $Z$ and $X$,
randomly to measure each signal after the PNS. }
\end{center}
\end{figure}

For the integrity, let us describe all the steps of this modified
Yan-Gao protocol as follows, including some steps same as those in
the original one \cite{YanGao}.

(M1) Same as the first step in the original Yan-Gao protocol,
Alice 1 prepares $nN$ qubits and each qubit is in one of the four
states $\{\vert 0\rangle, \vert 1\rangle, \vert +\rangle, \vert
-\rangle \}$ according to the bits in the two binary strings $A_1$
and $B_1$. She sends the qubits to Alice 2.

(M2) Similar to Alice 1, Alice 2 create two binary strings $A_2$
and $B_2$. For each of the $nN$ qubits, she performs the operation
$I$ ($U$) on it if the corresponding bit in $A_1$ is 0 (1).
Simultaneously she has to operate the qubit with $I$ or $H$
according to the corresponding bit in $B_2$ is 0 or 1,
respectively.  She sends the photons to Alice 3.

Certainly, before Alice 2 operates the $nN$ photons, she chooses
randomly a sufficiently enough subset of photons as the samples
for eavesdropping check. First, she splits each sample signal with
a PNS, and then measures each signal with the MB $Z$ or $X$
choosing randomly, shown in Fig.1. Moreover, she should analyze
the error rate $\varepsilon_s$ of the samples by means that she
requires Alice 1 to tell her the original states of the samples.
If the error rate is reasonably lower than the threshold
$\varepsilon_t$, Alice 2 continues the quantum communication to
next step, otherwise she aborts it.

(M3) Alice $i$ operates the qubits like Alice 2, $i=3,4,\ldots,
m$. For analyzing the error rate of the samples, she requires all
the members before her to tell her the original state or the
operations they chose.

(M4) Alice $m$ sends $N$ qubits to each member in group 2, Bob $j$
($j=1, 2, \ldots, n$) in turn. After they receive their qubits,
the members in group 1 publicly announce the strings $B_1, B_2,
\ldots, B_m$.

(M5) Bob $j$ measures each of his qubits with the MB $Z$ or $X$
according to the XOR (i.e., $\oplus$) results of corresponding
bits in the strings $B_1, B_2, \ldots, B_m$.

(M6) All members in group 1 complete the error rate analysis of
the transmission between the two groups. To this end, all Alice
require each of the member in group 2 to publish a subset of
results chosen randomly, and analyze the error rates of the
samples. If the channel is secure, the XOR results of Bob $j$'s
corresponding bits can be used as key bits for secret sharing,
otherwise they discard the results obtained and repeat the quantum
communication from the beginning.

In fact, this modified Yan-Gao protocol is secure with just a
little of modification of the original Yan-Gao protocol
\cite{YanGao}. That is, each of the members in group 1 performs an
eavesdropping check for the transmission of the qubits with a PNS
and two single-photon detectors. The principle of the
eavesdropping checks is the same as that in BB84 quantum key
distribution  protocol \cite{BB84,BB842}. So the transmission of
qubits between two authorized members in the two groups is secure.

This work is supported by the National Natural Science Foundation
of China under Grant Nos. 10447106, 10435020, 10254002, A0325401
and 10374010, Beijing Education Committee under Grant No.
XK100270454, the Hebei Natural Science Foundation of China under
Grant Nos. A2004000141 and A2005000140, and the Key Nature Science
Foundation of Hebei Normal University.

\end{document}